\documentclass[12pt]{iopart}
\usepackage{graphicx}

\begin{document}

\topical{Andreev reflection and order parameter symmetry in heavy-fermion superconductors: the case of CeCoIn$_5$}

\author{W K Park and L H Greene}

\address{Department of Physics and Frederick Seitz Materials Research Laboratory, University of Illinois at Urbana-Champaign, Urbana, Illinois 61801, USA}
\ead{wkpark@illinois.edu}

\begin{abstract}
We review the current status of Andreev reflection spectroscopy on the heavy fermions, mostly focusing on the case of CeCoIn$_5$, a heavy-fermion superconductor with a critical temperature of 2.3 K. This is a well-established technique to investigate superconducting order parameters via measurements of the differential conductance from nanoscale metallic junctions. Andreev reflection is clearly observed in CeCoIn$_5$ as in other heavy-fermion superconductors. Considering the large mismatch in Fermi velocities, this observation seemingly appears to disagree with the Blonder-Tinkham-Klapwijk (BTK) theory. The measured Andreev signal is highly reduced to the order of maximum $\sim$ 13\% compared to the theoretically predicted value (100\%). The background conductance exhibits a systematic evolution in its asymmetry over a wide temperature range from above the heavy fermion coherence temperature down to well below the superconducting transition temperature. Analysis of the conductance spectra using the extended BTK model provides a qualitative measure for the superconducting order parameter symmetry, which is determined to be $d_{x^2-y^2}$-wave in CeCoIn$_5$. It is found that existing models do not quantitatively account for the data, which we attribute to the intrinsic properties of the heavy fermions. A substantial body of experimental data and extensive theoretical analysis point to the existence of two fluid components in CeCoIn$_5$ and other heavy-fermion compounds. A phenomenological model is proposed employing a Fano interference effect between two conductance channels in order to explain both the conductance asymmetry and the reduced Andreev signal. This model appears plausible not only because it provides good fits to the data but also because it is highly likely that the electrical conduction occurs via two channels, one into the heavy electron liquid and the other into the conduction electron continuum. Further experimental and theoretical investigations will shed new light on the mechanism of how the coherent heavy-electron liquid emerges out of the Kondo lattice, a prototypical strongly correlated electron system. Unresolved issues and future directions are also discussed.
\end{abstract}

\submitto{\JPCM}
\maketitle

\section{Introduction}

It is well understood that the resistance minimum observed in metallic alloys diluted with small amount of magnetic impurities (e.g., Au-Fe or Cu-Fe) is caused by the spin-spin scattering between conduction electrons and local moments \cite{kondo64}. At temperatures approaching absolute zero, the magnetic moments of impurity atoms are screened by the conduction electron spins, prohibiting the divergence of the resistance due to the logarithmic scattering rate. This single impurity Kondo effect has been extensively studied in a variety of condensed matter systems \cite{gruner74,kouwenhoven01}. If localized magnetic moments form a dense ordered array, i.e., Kondo lattice, the underlying physics becomes much more complicated and its understanding has been a great challenge to theorists and experimentalists. According to Doniach's Kondo lattice model \cite{doniach77}, the fate of the local moments in the absence of direct dipolar interaction depends on the relative strength of two energy scales: one for Kondo interaction ($k_{\scriptsize \textrm{B}}T_{\scriptsize \textrm{K}} = De^{-1/2JN(0)}$); and the other for Ruderman-Kittel-Kasuya-Yoshida (RKKY) interaction ($k_{\scriptsize \textrm{B}}T_{\scriptsize \textrm{RKKY}} = J^2N(0)$), where $k_{\scriptsize \textrm{B}}$ is the Boltzmann constant, $D$ is the bandwidth, $J$ is the exchange coupling and $N$(0) is the conduction electron density of states (DOS). The ground state will be an antiferromagnetically ordered one if $T_{\scriptsize \textrm{RKKY}} > T_{\scriptsize \textrm{K}}$, whereas a Fermi liquid of heavy electrons emerges in the opposite case. This heavy electron liquid, also called a Kondo liquid, does not exhibit spontaneous magnetism. Many theoretical and experimental efforts have been made towards a microscopic understanding of such phase diagrams \cite{hewson93,coleman07}, particularly near quantum critical points \cite{stewart01,lohneysen07,gegenwart08}. The microscopic mechanism for the formation of heavy electron bands remains unknown as the topic of the most fundamental importance to this field. A key question is how the localized f-electrons embedded in a quantum sea of conduction electrons acquire the itinerancy over the system. One scenario is that the f-electrons become delocalized via hybridization with the conduction electrons: a highly dynamical and collective process. This collective compensation of Kondo lattice spins by conduction electron spins forms new Bloch states and results into a narrow renormalized band of heavy electrons \cite{coleman07}.

The majority of heavy electron materials are based on some rare earth (mostly Ce and Yb) or actinide (mostly U) elements that have valence electrons in the highly localized 4f or 5f orbitals. The overlap between neighboring f orbitals is negligible since they are highly confined to the nuclei over a length scale less than the interatomic spacing. Among dozens of Ce-based heavy-fermion compounds discovered so far, the 1-1-5 family CeMIn$_5$ (M = Co, Rh, Ir) have added to new avenues for the study of rich and novel physics including quantum criticality, the Fulde-Ferrell-Larkin-Ovchinnikov (FFLO) phase transition and the interplay between superconductivity and magnetism (see \cite{thompson03,sarrao07} and references therein). The Ce 1-1-5 family is a subgroup of a larger class: Ce$_n$M$_m$In$_{3n+2m}$. Here, the Ce atoms occupy basal planes of a tetragonal crystal structure. Measurements of field-oscillatory magnetization (de Haas-van Alphen effect) in CeCoIn$_5$ have identified multiple bands crossing the Fermi level \cite{settai01,shishido02}. Major Fermi surfaces consist of warped cylinders for heavy electron and hole bands and small ellipsoidal pockets for light holes \cite{settai01,shishido02}. The resistivity maximum that signifies the beginning of heavy fermion coherence ($T$*) occurs around 45 K and a rapid drop is followed with decreasing temperature \cite{petrovic01}. The emerging heavy electron liquid is found to disobey the Landau theory of Fermi liquid as evidenced by the $T$-linear (below $\sim$ 20 K) temperature dependence of the resistivity and the ln$T$ dependence of the electronic specific heat coefficient, to name a few \cite{thompson03}. This non-Fermi liquid behaviour implies proximity to a quantum critical point \cite{sidorov02,paglione03,bianchi03}. Superconductivity in CeCoIn$_5$ sets in below 2.3 K and the large specific heat jump at $T_{\scriptsize \textrm{c}}$ indicates that heavy electrons participate in the Cooper pairing \cite{petrovic01}.

Most of known heavy-fermion superconductors exhibit unconventional pairing symmetry and gap structure \cite{coleman07}. For CeCoIn$_5$, the power law dependences of electronic and heat transport and the thermodynamic properties point to the existence of line nodes on the Fermi surface \cite{thompson03}. Magnetic field-angle dependent thermal conductivity \cite{izawa01} and specific heat \cite{aoki04} measurements exhibit four-fold oscillations in the basal plane, further supporting that the gap has $d$-wave symmetry \cite{izawa01,aoki04,matsuda06}. However, these results are seemingly contradictory for the precise node locations. Thermal conductivity indicates $d_{x^2-y^2}$ \cite{izawa01} and specific heat supports $d_{xy}$ \cite{aoki04}. There have been theoretical analyses which explain the controversy \cite{vorontsov06}. Not only is the interpretation of these experiments complex \cite{vorontsov06}, but also they intrinsically cannot provide phase information of the order parameter. An unambiguous determination of the pairing state between $d_{xy}$ and $d_{x^2-y^2}$ is of crucial importance to elucidating the pairing mechanism (e.g., see \cite{scalapino87}).

Tunneling spectroscopy as a probe of the superconducting order parameter has proven extremely useful and thus has been extensively adopted for both conventional \cite{wolf85} and unconventional \cite{zasadzinski03,fischer07} superconductors. This technique was instrumental in confirming the predictions of the Bardeen-Cooper-Schrieffer (BCS) theory of superconductivity in conventional systems by measuring the quasiparticle DOS and phonon modes mediating the Cooper pairing (\cite{wolf85} and references therein). However, such a direct measurement of the order parameters in heavy-fermion superconductors has proven much more difficult compared to conventional superconductors and even to other unconventional superconductors, where planar tunneling and scanning tunneling spectroscopy (STS) have played key roles. This is due to their unique materials properties such as extremely low $T_{\scriptsize \textrm{c}}$, short coherence length, high reactivity, dependence of superconducting properties on the structural disorder, difficulty of thin film growth and non-cleavability. As an alternative technique, Andreev reflection (AR) measurements based on point-contact methods have been frequently adopted \cite{lohneysen96,naidyuk98,naidyuk05}. This technique, called point-contact spectroscopy (PCS), has been successful to some extent but also has raised some issues including non-spectroscopic effects and how AR is possible at a normal-metal/heavy-fermion superconductor interface. The non-spectroscopic effects can originate from a contact being in the non-ballistic regime due to short mean free path, inter-grain Josephson coupling, etc, as will be discussed below.

We review the current status of experimental and theoretical studies of AR in heavy-fermion systems, focusing on CeCoIn$_5$. In section 2, basic theories on the charge transport in normal-metal/superconductor (N/S) junctions are reviewed. Basics of AR measurements using PCS technique are described in section 3 and technical issues in PCS are discussed in sections 4. Section 5 is devoted to the discussion whether an observation of a zero-bias conductance peak can be taken as a signature of the sign change in the order parameter. In section 6, our experimental results are presented in terms of the superconducting order parameter symmetry. How AR is possible in heavy fermions is discussed in section 7. Possible origins of the conductance asymmetry, a common feature in all our measurements on the Ce 1-1-5 family, are addressed in section 8. Our conductance modeling based on a possible Fano resonance in the Kondo lattice is described in section 9. Unresolved issues and future directions are presented in section 10. Finally, conclusive remarks will follow.

\section{Charge transport across an N/S interface}

\begin{figure}[t]
\begin{center}
\includegraphics[scale=0.7]{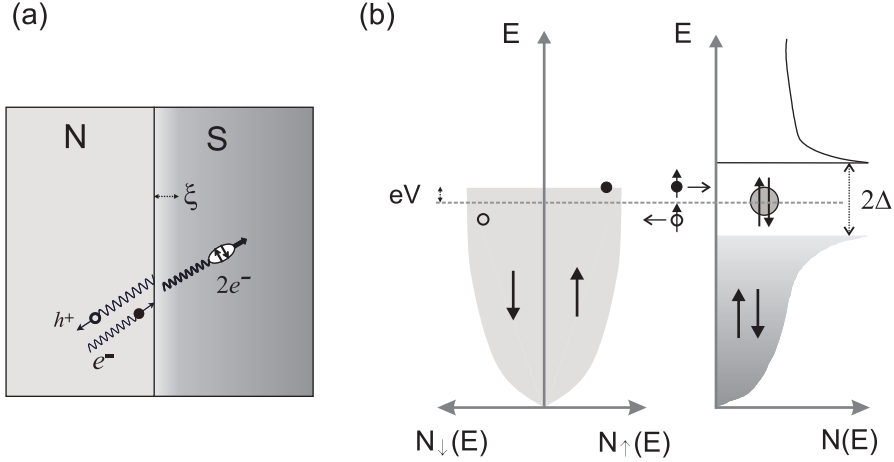}
\end{center}
\caption{\label{fig:AR} A schematic of Andreev reflection process. (a) A real space picture. N and S denote a normal-metal and superconducting electrode, respectively. (b) A picture in the E (energy) vs. DOS space when the S is biased positively. The dashed line indicates the Fermi level. An electron incoming from the N side, impinging on the N/S interface, is retroreflected as a hole and get paired with another electron with opposite spin and momentum, transferring as a Cooper pair into the superconductor over a region $\xi$ (superconducting coherence length) near the interface. Note that energy, momentum, spin and charge (including a Cooper pair) are conserved in this process.}
\end{figure}

While trying to explain an unexpected increase of thermal resistance in the intermediate state of type-I superconductors, in 1964 Andreev \cite{andreev64} discovered that there must be an additional scattering at interfaces between normal and superconducting domains. Its microscopic process was precisely derived by solving the boundary condition problem using Bogoliubov-de Gennes equations. The hidden scattering is retroreflection of an electron as a hole as it approaches an N/S interface from the N side. This particle-hole conversion process, called Andreev reflection, is essentially a scattering off a superconducting pair potential, during which energy, momentum and spin are conserved. The missing electronic charge is transferred as a Cooper pair into the superconductor, doubling the conductance within the energy gap, $\Delta$. A schematic of this process is depicted in figure~\ref{fig:AR}. If the normal metal is replaced with a ferromagnet, AR is suppressed since the DOS for the opposite spin band is severely reduced (see figure~\ref{fig:AR}(b) with unbalanced spin bands in N). This phenomenon has been extensively utilized to measure the spin polarization of ferromagnetic materials \cite{soulen98}. 

At about the same time as Andreev formulated the occurrence of AR, de Gennes and Saint-James \cite{degennes63} considered a similar boundary condition problem in an N/S bilayer system and discovered that the quasiparticle DOS is modified by the induced bound states. Although such AR-related effects had been known for a long time (e.g., \cite{rowell73}), AR was not considered as a direct measure of the superconducting energy gap until a theory came out in 1982. While studying the charge transport properties of N/S and S/N/S junctions, Blonder, Tinkham and Klapwijk (BTK) came up with an elegant theoretical formulation to explain various characteristics in a unified model \cite{btk82,blonder82}. They considered a boundary condition problem with a delta-function potential barrier at an N/S interface, $H\delta(x)$, and the quasiparticle energy in the superconductor determined by the Bogoliubov-de Gennes equations. For an incoming electron impinging on the N/S interface from the N side, there are four possible trajectories: normal reflection, AR as a hole, transmission into the same branch and transmission with a branch crossing. In the BTK theory, the differential conductance across an N/S interface is given by the following formula.

\begin{equation}
\frac{dI}{dV}(V)= 2N(0)e v_{\scriptsize \textrm{F}} S \int_{-\infty}^{\infty}dE \frac{\partial f(E-eV)}{\partial (eV)}\left[ 1+A(E)-B(E) \right],
\label{eq:btk}
\end{equation}
where $N(0)$ is the DOS at the Fermi level, $e$ is the electronic charge, $v_{\scriptsize \textrm{F}}$ is the Fermi velocity, $S$ is the junction area, $f$ is the Fermi distribution function, $A(E)$ is the AR probability and $B(E)$ is the normal reflection probability.

\begin{figure}[t]
\begin{center}
\includegraphics[scale=1.2]{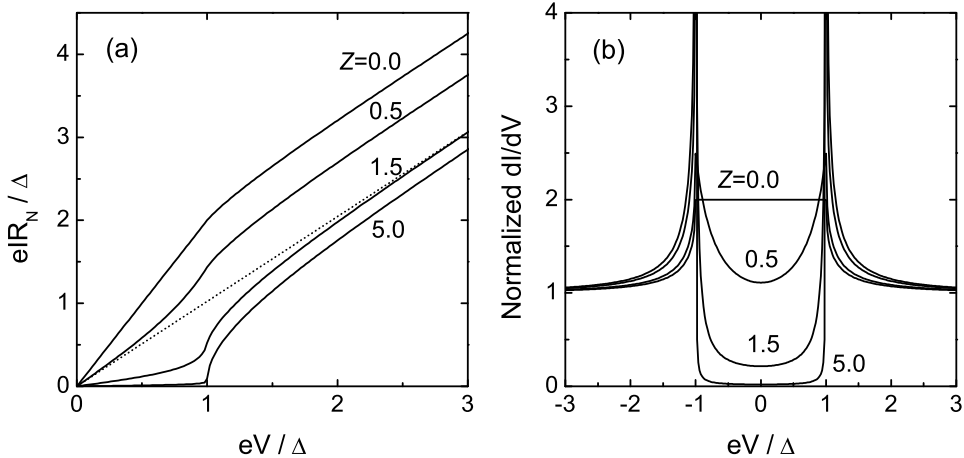}
\end{center}
\caption{\label{fig:sBTK} (a) Current vs. voltage and (b) normalized dI/dV vs. voltage characteristics calculated by the BTK formula for a normal-metal/s-wave superconductor junction at $T$=0. The dotted line in (a) is an asymptote to the $Z$=1.5 curve. $R_{\scriptsize \textrm{N}} = R_0(1+Z^2)$, where $R_0$ is the normal state resistance of a purely metallic junction. In the AR regime, an excess current is observed, which is the intercept obtained when the high-bias part of an I-V curve is extrapolated to the y-axis. Note that the BTK theory reproduces the transitional behaviour from AR to tunneling with a single parameter, $Z$, the dimensionless barrier strength.}
\end{figure}

A remarkable aspect of this model is that it can reproduce characteristic conductance features from AR to tunneling as a function of single parameter, $Z \equiv H/\hbar v_{\scriptsize \textrm{F}}$, the dimensionless barrier strength. This is demonstrated in figure~\ref{fig:sBTK} for an $s$-wave superconductor as calculated by (\ref{eq:btk}). For a purely metallic junction ($Z=0$), the subgap conductance is doubled with respect to the normal state value, which is due to 100\% probability of AR for $E < \Delta$. With increasing $Z$, a double-peak structure develops and the conductance at zero bias decreases, both features due to the energy dependence of AR probability. For a large $Z$ value, say, $Z > 3$, AR is almost suppressed and the conductance shape is dominated by single particle tunneling. In this limit, the tunneling conductance maps out the quasiparticle DOS which, in the strong coupling case, may contain information on the bosonic modes mediating the Cooper pairing \cite{mcmillan65}. Since AR is a scattering off a pair potential, it is sensitive to the phase coherency of the Cooper pairs \cite{beenakker99,pannetier00}. Thus, AR measurements on unconventional superconductors may give different information than tunneling spectroscopy. This is an important issue that has been debated actively in the case of high-$T_{\scriptsize \textrm{c}}$ cuprate superconductors \cite{deutscher99,hufner08}.

In an effort to analyze the conductance data obtained from Nb/Cu point-contact junctions, Blonder and Tinkham introduced a second term into the barrier strength parameter as follows \cite{bt83}.
\begin{equation}
Z_{\scriptsize \textrm{eff}} = [Z^2 + (1-r)^2/4r]^{1/2},
\label{eq:barrierZ}
\end{equation}
where $r \equiv v_{\scriptsize \textrm{F}}^{\scriptsize \textrm{N}} /v_{\scriptsize \textrm{F}}^{\scriptsize \textrm{S}}$, the ratio of Fermi velocities in the normal-metal ($v_{\scriptsize \textrm{F}}^{\scriptsize \textrm{N}}$) and superconducting ($v_{\scriptsize \textrm{F}}^{\scriptsize \textrm{S}}$) electrodes. They observed that the fitting parameter $Z_{\scriptsize \textrm{eff}}$ ranged $0.3 - 1$ and interpreted that the minimum value of 0.3 is consistent with the $Z_{\scriptsize \textrm{eff}}$ value calculated by (\ref{eq:barrierZ}) using $r \approx 1.75$ for Nb/Cu and assuming no insulating barrier ($Z$=0). Our experimental data on dozens of Nb/Au ($r \approx 1.0$ \cite{ashcroft76}) point-contact junctions over a range of sample-tip pressures have shown that $Z_{\scriptsize \textrm{eff}} \ge 0.2$. This indicates that it may not be justified to extract the Fermi velocity mismatch contribution from the obtained $Z_{\scriptsize \textrm{eff}}$ values, contrary to their interpretation and to the claim by Deutscher and Nozi{\'e}res \cite{deutscher94} that one can determine the mass enhancement factor by comparing velocities obtained from PCS with known values. Regarding this issue, Lukic considered generalized boundary conditions in the BTK problem and argued that the effective mass is just a parameterization of the unknown microscopic parameters \cite{lukic}.

In heavy-fermion materials, the electronic mass is highly enhanced (by an order of $\sim 10-10^3$) with a correspondingly reduced Fermi velocity. According to (\ref{eq:barrierZ}), a junction with a heavy fermion as one electrode and a conventional metal as the other is inherently in the tunneling regime (e.g., $Z_{\scriptsize \textrm{eff}} \ge 5$ for Au/CeCoIn$_5$) and thus AR cannot occur. However, AR has been frequently observed in such junctions as we present our data below and as observed by others (\cite{naidyuk98} and references therein). Deutscher and Nozi{\'e}res addressed this discrepancy by proposing that the boundary conditions are not affected by the mass enhancement factor \cite{deutscher94}. These issues are further discussed in section 7.

\section{Basics of point-contact spectroscopy}

Detailed reviews on this topic have been published, e.g., \cite{naidyuk05}. Here we only present fundamental concepts and elements of PCS. Let us begin by considering electrical conduction across a metallic contact between two identical metal electrodes. Assume the two metals are connected only through a constriction of radius $a$ and otherwise electrically isolated. If a bias voltage is applied across the junction, the resulting conductance will depend on how the electric field distributes in the contact region. Here, the most relevant length scales are the electronic mean free paths, elastic ($l$) and inelastic ($l_{\scriptsize \textrm{in}}$), in comparison to the contact radius $a$. If $2a \ll l, l_{\scriptsize \textrm{in}}$, electrons gain or lose energy only when crossing the interface and their velocity change is proportional to the applied voltage. This gives rise to an ohmic current-voltage characteristic and the contact resistance is given by the following formula, known as Sharvin resistance \cite{sharvin65}:
\begin{equation}
R_{\scriptsize \textrm{S}} = \frac{4\rho l}{3\pi a^2},
\label{eq:sharvin}
\end{equation}
where $\rho$ is the resistivity of the metal. This limit ($2a \ll l, l_{\scriptsize \textrm{in}}$) is called ballistic or Sharvin limit. Note that the resistance depends only on the contact area since the product $\rho l$ is constant in the simplest Drude picture. In the opposite limit, namely, if $2a \gg l, l_{\scriptsize \textrm{in}}$, the electron distribution would be that as in the bulk. Then, the point-contact resistance is given by the formula for a bulk sample of length $2a$ and diameter $2d$ \cite{holm67}, except for a geometric factor:
\begin{equation}
R_{\scriptsize \textrm{M}} = \frac{\rho}{2a}.
\label{eq:maxwell}
\end{equation}
This limit ($2a \gg l, l_{\scriptsize \textrm{in}}$) is called the thermal or Maxwell limit and a local Joule heating can predominate the electrical transport process.

\begin{figure}[b]
\begin{center}
\includegraphics[scale=0.7]{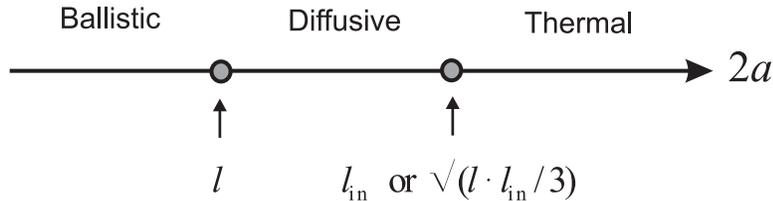}
\end{center}
\caption{\label{fig:PCSregime} Schematic of the PCS regimes. $a$ is the radius of a point contact and $l$ ($l_{\scriptsize \textrm{in}}$) is the elastic (inelastic) mean free path of electrons in the electrodes. Ballistic: $2a \ll l, l_{\scriptsize \textrm{in}}$. Diffusive: $l \ll 2a \ll \sqrt{l \cdot l_{\scriptsize \textrm{in}} / 3}$. Thermal: $2a \gg l, l_{\scriptsize \textrm{in}}$.}
\end{figure}

By solving Boltzmann transport equations for a point-contact junction with an arbitrary value for $K \equiv l/a$, Wexler derived a generalized formula for the point-contact resistance as follows \cite{wexler66}.
\begin{equation}
R_{\scriptsize \textrm{PC}} = \frac{4\rho l}{3\pi a^2} \left[ 1 + \frac{3\pi}{8K} \gamma(K) \right],
\label{eq:wexler}
\end{equation}
where $\gamma(K)$ is a smooth function of $K$. The intermediate region between ballistic and thermal regimes is called the diffusive regime, where $l \ll 2a \ll \sqrt{l \cdot l_{\scriptsize \textrm{in}} / 3}$ and electrons undergo multiple elastic scattering across the constriction. In the case of an N/S point-contact junction, this may reduce the AR signal. In figure~\ref{fig:PCSregime}, the three regimes are indicated in relation to the contact size and mean free paths. An extreme limit, not relevant to the PCS discussed in this paper, can occur if the contact size becomes comparable to the Fermi wavelength, in which case quantum size effects dominate \cite{srikanth92,agrait03}. This regime is reached in atomic size quantum point contacts \cite{agrait03}.

If a point-contact junction is in the ballistic limit, energy-dependent quasiparticle scatterings appear as nonlinearities in the current-voltage characteristics, analysis of which gives spectroscopic information of the material under study. What makes this PCS possible is that the contact area remains cold due to a large mean free path compared with the contact size and any energy dissipation occurs away from the junction. In contrast, in the thermal regime, energy dissipation occurs within the contact volume, causing local heating. This PCS technique was first demonstrated by Yanson in 1974 in his second harmonic measurements of micro-shorted thin film tunnel junctions \cite{yanson74}, from which he obtained phonon spectrum in Pb, similar to that obtained in the tunneling measurements by McMillan and Rowell \cite{mcmillan65}. Shortly after, Jansen and coworkers developed a more controllable method to make point-contact junctions \cite{jansen77}. This technique, called the `spear-anvil' method, has been adopted most widely for PCS because of its simplicity. During its early days, PCS was mostly utilized to investigate phonon spectra in metals and alloys \cite{jansen80,duif89}. Since the BTK theory came out \cite{btk82} to explain the experimental data nicely \cite{blonder82,bt83}, PCS has been frequently adopted as an alternative tool to planar tunneling or STS for the study of superconducting gap structures of many novel and unconventional superconductors owing to its versatility and technical simplicity \cite{naidyuk98,naidyuk05,deutscher05}. In the majority of PCS experiments on superconductors, the differential conductance vs. bias voltage data have been analyzed successfully using the BTK model to determine their superconducting order parameters.

There exist many reports in the literature where the PCS data may not necessarily be spectroscopic. As discussed above, for purely spectroscopic measurements, the junction must be in the ballistic regime. A first check of this is to estimate junction size using (\ref{eq:wexler}) and compare with known values for the mean free paths . Because in general point-contact junctions are made by bringing two metals into mechanical contact on the nanoscale, data can also be affected by factors other than the nominal contact size, such as contact pressure \cite{gloos95} and geometry \cite{park06prl}, meaning that the nominal contact size being in the ballistic limit alone cannot guarantee the spectroscopic nature of the data. The most  important diagnostic is the reproducibility and consistency of the conductance spectra point to point, sample to sample and along different crystallographic orientations, as we discuss in the next three sections.

\section{Experimental methods}

Like for other class of superconductors, the spear-anvil method has been employed widely for the PCS investigation of heavy-fermion superconductors. Their much lower $T_{\scriptsize \textrm{c}}$'s (typically less than 1-2 K) make its implementation complicated because of the inevitable involvement of a $^3$He or dilution refrigerator. In these cases, a purely mechanical approach of the tip to the sample based on the differential screw mechanism, which is widely adopted for $^4$He systems, is not suitable. Instead, a combination of mechanical and piezoelectric adjustments is more common, where a coarse approach is made by a fine-pitched screw and a fine adjustment is made by driving piezoelectric elements \cite{park06rsi}. Although mechanically cut wires (typically Au or Pt-Ir) have been used, electrochemically polished tips provide better control over the tip size and surface morphology. This is an important technical issue since it could affect the contact geometry and thus the conductance spectra. 

It is also crucial to prepare smooth and clean sample surfaces in order to obtain spectroscopic data. Almost all materials investigated by PCS have surface layers that are not the same as in bulk. These layers could be precipitates from growth, surface reconstruction or due to oxide growth when exposed to air. In some cases, these layers can be penetrated by the metal tip during the tip approach, enabling one to probe bulk characteristics. Otherwise, the surface layer can act as a potential barrier, proximity or degraded layer. If the surface layer is thin, the potential barrier case will still provide spectroscopic data which can be analyzed by the BTK theory, but approaching, or in, the tunneling limit. In the other two cases, obtaining spectroscopic data is more difficult, particularly if the surface layer is thick on the order of the coherence length or longer.

It is ideal if a superconductor under study can be prepared such that it has flat and smooth surfaces along all major crystallographic orientations. Most unconventional superconductors are first produced in bulk (poly- or single-crystalline) forms. PCS on polycrystals can reveal some information on the gap values but the momentum direction is not identified or defined well, meaning that data analysis and interpretation might not provide crystallographic orientation-dependent information. In addition, because of a granular structure, measured data are vulnerable to artificial effects such as intergrain Josephson coupling \cite{shan03}. Thus, single crystalline samples are highly desirable. A drawback, particularly in anisotropic materials, is that crystals are often grown as platelets only along the preferred surface orientations. Thin film samples can be grown along desired directions by adjusting growth conditions but heavy-fermion compounds are known to be extremely difficult to synthesize in thin film form. For PCS, we rely on the high-quality single crystals that have been grown and the preferred as-grown surface of a single crystal is first investigated. For measurements along the other directions, broken edges are frequently used. Preparation of surfaces by polishing is rarely adopted due to the possibility of surface degradation during the polishing process. However, this approach for sample preparation has been successful for the AR study of CeCoIn$_5$ \cite{park07physc}.

The heavy-fermion compound CeCoIn$_5$ has a tetragonal crystal structure with the Ce atoms occupying the basal plane. Many experimental reports have pointed to a $d$-wave symmetry of the Cooper pairing state \cite{thompson03,izawa01,aoki04,matsuda06}. Considering its known Fermi surfaces \cite{settai01,shishido02}, the gap nodes are supposed to form lines along the $c$-axis, similarly to the case of cuprate superconductors. From the group theoretical point of view on the possible pairing state, there is a correspondence between real and momentum space regarding the gap node directions in $d$-wave superconductors \cite{annett96}. Therefore, it is essential to make PCS measurements into the $ab$ plane to probe along nodal and antinodal directions. More specifically, measurements into (100) and (110) surfaces should be made to identify the location of nodes. The largest face (typically ranges a few mm) of an as-grown CeCoIn$_5$ crystal is perpendicular to the $c$-axis. Thus, this surface is naturally chosen for PCS along the $c$-axis. Since the thickness of single crystals ranges only a few hundred $\mu m$'s, it is necessary to devise a method to hold a sample securely such that its in-plane direction is aligned with the tip axis. This can be achieved by embedding a crystal into a mold made of low-temperature epoxy and cutting and polishing it \cite{park07physc}. 

CeCoIn$_5$ crystals are grown by the flux method using excess indium flux \cite{petrovic01}. Because of the nature of this method, samples can contain indium precipitates, which may affect the PCS measurements. To eliminate this possibility, one slightly etches crystals using hydrochloric acid, noting that CeCoIn$_5$ is also dissolved although at much slower rate than indium. Microscopically rough surfaces can therefore be produced, causing spurious effects. Again, to ensure the spectroscopic nature of the conductance spectra, reproducibility is the most important diagnostic. In particular, data should be compared along all major crystallographic orientations. In general, considering the extreme cleanness of CeCoIn$_5$ (the electronic mean free path ranges several $\sim \mu m$'s at low temperature \cite{kasahara05}), it is highly feasible to perform PCS measurements in the ballistic regime.   

\section{Zero-bias conductance peak vs. Andreev bound states}

Three independent groups have reported results of PCS on CeCoIn$_5$. Goll \textit{et al}. \cite{goll03} first reported two different types of conductance spectra obtained from point contacts along the $c$-axis: one with double-peak and zero-bias dip structure, and the other with a zero-bias peak and surrounding dips. The latter feature was interpreted as hinting at an unconventional symmetry ($d$-wave was implied). Rourke \textit{et al}. \cite{rourke05} also reported two different types of spectra from nominally $c$-axis contacts: one with a zero-bias peak and hump structure, and the other with a double hump structure. These authors claimed both $d$-wave symmetry and multiple order parameter components. The validity of the claim for $d$-wave symmetry based simply on the observation of a zero-bias conductance peak (ZBCP) has been questioned \cite{park06prl,sheet06} because many other origins for a ZBCP are also known.

\begin{figure}[t]
\begin{center}
\includegraphics[scale=1.2]{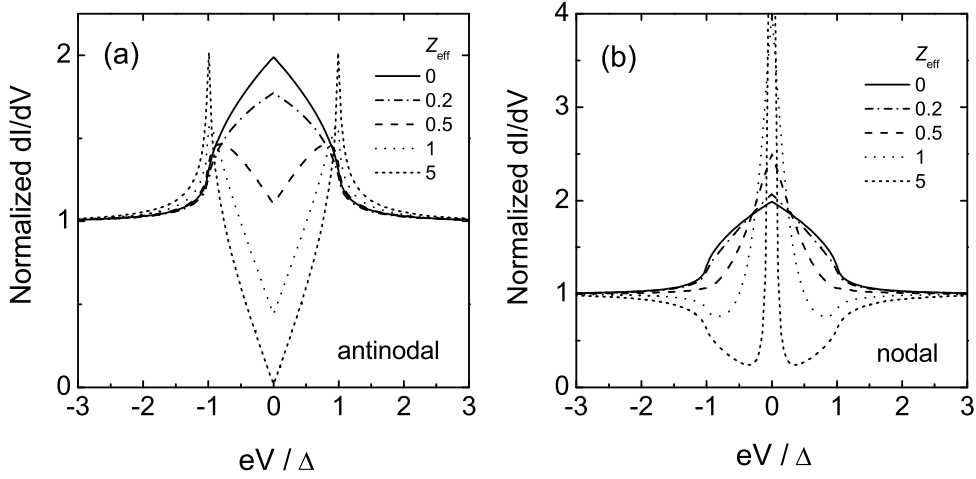}
\end{center}
\caption{\label{fig:dBTK} Normalized dI/dV vs. voltage characteristics calculated by the extended BTK formula for a normal-metal/$d$-wave superconductor junction at $T$=0. (a) The junction normal is along the antinodal direction. The conductance behaviour is qualitatively similar to the $s$-wave case except for the cusp-like (AR) and V-shape (tunneling) features near zero bias, which are due to a $d$-wave superconducting DOS. (b) The junction normal is along the nodal direction. Completely different behaviours are observed due to the surface Andreev bound states.}
\end{figure}

Figure~\ref{fig:dBTK} shows normalized dI/dV curves calculated for a normal-metal/$d$-wave superconductor junction using the kernel developed by Tanaka and Kashiwaya {\it et al.}   \cite{tanaka95,kashiwaya96,kashiwayatanaka} as a function of $Z_{\scriptsize \textrm{eff}}$, the effective barrier strength in (\ref{eq:barrierZ}). Here, it is assumed that the injected electrons have a momentum distribution over the whole range of angle (90 $\deg.$) around the junction normal. In practice, in the tunneling limit, tunneling electrons may have a much narrower momentum distribution, known as tunneling cone effect \cite{beuermann81}. In the limit of a 90 degree tunneling cone, there is no difference between the $c$-axis, nodal or antinodal directions as there is an integration over the full half space of momentum, as shown in figures~\ref{fig:dBTK}. For an antinodal or $c$-axis junction, the cusp-like feature and linear slope near zero bias for purely metallic ($Z_{\scriptsize \textrm{eff}} = 0$) and tunnel junctions ($Z_{\scriptsize \textrm{eff}} = 5$), respectively, exhibit the characteristic DOS of a $d$-wave superconductor. This is just a qualitative difference from the $s$-wave case, where a flat conductance shape is seen due to a fully isotropic opening of the gap. As $Z_{\scriptsize \textrm{eff}}$ is increased from zero, a double-peak and zero-bias dip structure develops, which is also similar to the $s$-wave case (see figure~\ref{fig:sBTK}(b)). An interesting case is the nodal direction with a finite $Z_{\scriptsize \textrm{eff}}$. The conductance shape varies in a strikingly different way, as displayed in figure~\ref{fig:dBTK}(b). Beginning with the same shape for $Z_{\scriptsize \textrm{eff}} = 0$ as for the antinodal junction, the conductance curve becomes narrower and sharper with increasing $Z_{\scriptsize \textrm{eff}}$, forming a ZBCP instead of a dip. This ZBCP originates from Andreev bound states (ABS) formed on the nodal surface due to a sign change of the $d$-wave order parameter around the Fermi surface \cite{hu94}. The ZBCP in tunnel junctions on hole-doped cuprate superconductors, well known to be $d$-wave, has been frequently observed \cite{deutscher05,covington97} and well understood theoretically \cite{fogelstrom97}. The ABS occurs due to the constructive interference between an incoming electron and an Andreev-reflected hole and can be formed at zero energy on a nodal surface of a $d$-wave superconductor because of the additional phase shift of $\pi$ \cite{hu94,lofwander01}. The same shape for a nodal junction as for an antinodal junction when $Z_{\scriptsize \textrm{eff}} = 0$ can be understood as due to smearing of ABS in the absence of barrier \cite{kashiwaya96,kashiwayatanaka}.

It has been frequently claimed that a ZBCP observed from PCS is a signature of ABS, thus, evidence for a sign change of the superconducting order parameter. However, as pointed out by us \cite{park06prl}, and Sheet and Raychaudhuri \cite{sheet06}, just an observation of a ZBCP does not corroborate such claims. A ZBCP can also arise from other physical mechanisms without requiring a sign change such as local heating \cite{gloos96}, intergrain Josephson coupling \cite{shan03}, etc. Therefore, it is essential to perform diagnostic measurements and analyses in order to prove that a ZBCP is a real signature of ABS. In the high-$T_{\scriptsize \textrm{c}}$ cuprate tunnel junctions, it is well established that ABS-originated ZBCPs are split either spontaneously or by applied magnetic field \cite{deutscher05,covington97}. One of the broadly accepted explanations is the Doppler shift of the bound state energy due to coupling to the superfluid momentum \cite{fogelstrom97}. The amplitude of an ABS-originated ZBCP is predicted to follow 1/$T$ dependence, in contrast with the ln$T$ behaviour of Kondo scattering effect in tunnel junctions. In the case of metallic junctions as in PCS, it is proposed that the critical magnetic field over which the ZBCP can be split is much stronger \cite{tanaka02}. Another important diagnostic measurement, particularly necessary for metallic junctions because the splitting may not be observed for the reasons above is to compare conductance spectra along different crystallographic directions. In the case of CeCoIn$_5$, given that $d$-wave symmetry is most likely, measurements along both (100) and (110) directions are essential to differentiate between $d_{x^2-y^2}$- and $d_{xy}$-wave symmetry.

Neither of the reports by Goll {\it et al.} \cite{goll03} and by Rourke {\it et al.} provided such diagnostic measurements, so origins of their observed ZBCPs remain unclear. Rourke {\it et al.}'s claim on the existence of multiple order parameters relies on their observation of multiple hump structures in the differential conductance data. However, such multiple hump or peak and dip structures have been sometimes observed in PCS study of a known single gap superconductor. To prove that the hump positions can be taken as gap values for multiple order parameters, diagnostic measurements including crystallographic orientation and temperature dependences should be made along with reproducibility. In general, considering the band structure and Fermi surface topology \cite{settai01,shishido02}, it is possible that the superconductivity in CeCoIn$_5$ has multiband nature, but it is a different issue to prove that measured PCS data actually reflect such features.

\section{Spectroscopic evidence for $d_{x^2-y^2}$ symmetry}

\begin{figure}[t]
\begin{center}
\includegraphics[scale=1.1]{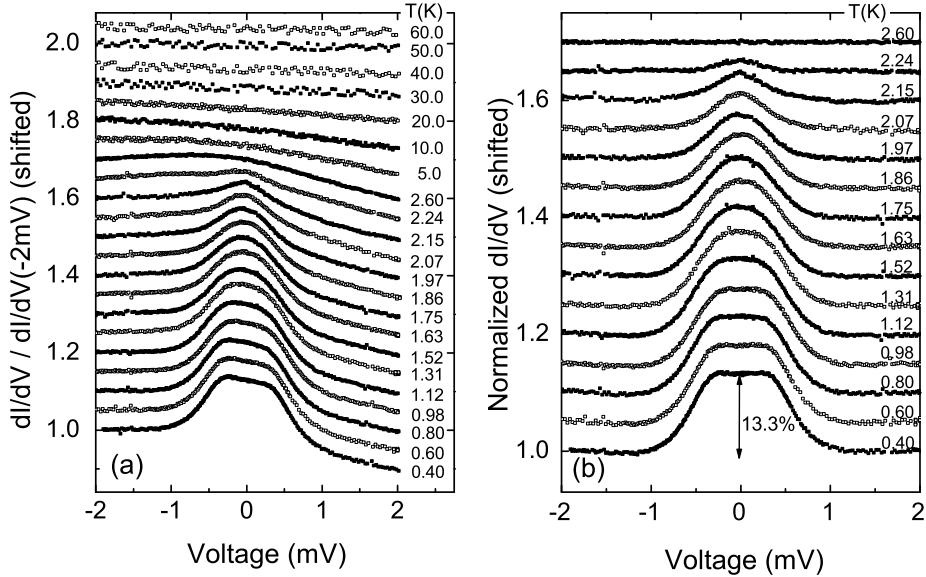}
\end{center}
\caption{\label{fig:PCS001} (a) Differential conductance spectra of a (001) CeCoIn$_5$ point-contact junction. Note the systematic evolution of the background conductance asymmetry and the enhancement of conductance near zero bias due to Andreev reflection. (b) Conductance curves normalized by the 2.6 K data. There is no dip in the raw data in (a), so the shallow dips around $\pm$ 1 mV at 400 mK in (b) are due to the normalization, not caused by local heating. The magnitude of AR conductance is on the order of $\sim$ 13\%. After \cite{park05}.}
\end{figure}

We have reported differential conductance spectra from Au/CeCoIn$_5$ point-contact junctions along all three crystallographic directions \cite{park07physc,park05,park08prl}. Figure~\ref{fig:PCS001}(a) displays temperature evolution of the conductance along the $c$-axis over a wide temperature range. At high temperatures, the conductance curves are symmetric and flat, which is characteristic of simple metallic junctions over a small bias region. With decreasing temperature, an asymmetry in the dI/dV curves is developed. This conductance asymmetry appears to begin below the heavy fermion coherence temperature $T^*$ ($\sim$ 45 K) \cite{nakatsuji04} and increases with decreasing temperature down to $T_{\scriptsize \textrm{c}}$ (2.3 K), below which it remains constant, as shown in figure~\ref{fig:PCS001}(a). As $T_{\scriptsize \textrm{c}}$ is crossed, the conductance near zero-bias begins to be enhanced and this enhancement increases with decreasing temperature. In the literature, it has been frequently adopted to analyze conductance data by normalizing them with respect to those just above $T_{\scriptsize \textrm{c}}$. Noting that our data below $T_{\scriptsize \textrm{c}}$ exhibit nearly the same background shape, we normalized them against the data taken at 2.6 K. The result is plotted in figure~\ref{fig:PCS001}(b). The temperature and bias voltage dependences and the shape of the conductance curves (BTK-like) near zero bias indicate the enhancement is due to AR. However, the magnitude of the AR conductance ($\sim$ 13\%) is greatly suppressed in comparison to the theoretical prediction (100\%) \cite{btk82}.

\begin{figure}
\begin{center}
\includegraphics[scale=1.2]{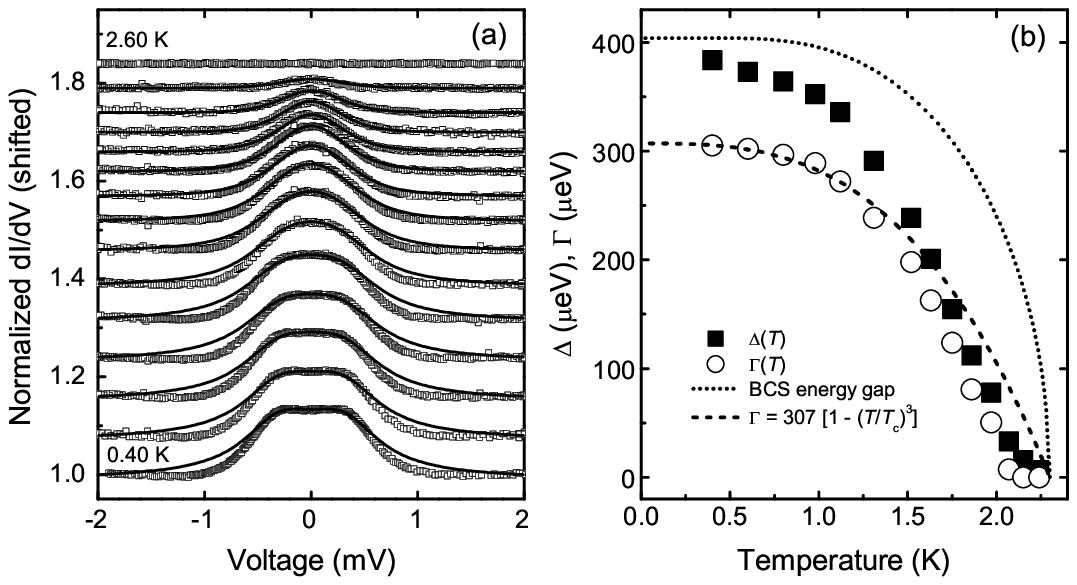}
\end{center}
\caption{\label{fig:sBTKfit} (a) Normalized dI/dV ($\opensquare$) of a (001) CeCoIn$_5$ junction in figure~\ref{fig:PCS001}(b) and best fit curves ($\full$) by the $s$-wave BTK model. (b) Energy gap ($\Delta$) and quasiparticle lifetime broadening factor ($\Gamma$) obtained from the $s$-wave fit. $\Delta$ predicted by the BCS theory ($\dotted$) and $\Gamma$ ($\dashed$) obtained from the zero-bias conductance fit. Note that the temperature dependence of $\Gamma$ is unphysical. After \cite{park05}.}
\end{figure}

Our attempt to fit the normalized conductance spectra on the (001) CeCoIn$_5$ junction using the extended BTK model by Tanaka and Kashiwaya {\it et al.} \cite{tanaka95,kashiwaya96,kashiwayatanaka} revealed that this model could not fully account for the data \cite{park05}. The best-fit curves deviate from the data substantially (figure~\ref{fig:sBTKfit}(a)) and the temperature dependence of the fitting parameter $\Gamma$, smearing factor \cite{dynes78}, basically tracks that of the gap, which is unphysical (figure~\ref{fig:sBTKfit}(b)) \cite{park05}. The large deviation around gap edges implies that the failure is not solely due to the large suppression of AR but also due to a large shift of the spectral weight \cite{park05}. This situation remains unchanged no matter which modified BTK model we adopt for $s$- or $d$-wave symmetry considering the Fermi surface mismatch in the two electrodes and the possible breakdown of Andreev approximation \cite{park05,park05spie,elenewski05,lukic05,mortensen99,golubov00}. All these results point to a necessity to develop a new theoretical framework to understand how AR is possible at all and why it is observed so reduced in heavy-fermion/conventional-metal interfaces. Even without the full theoretical framework, PCS can be used as a spectroscopic probe of the order parameter symmetry as we discuss below.

\begin{figure}[t]
\begin{center}
\includegraphics[scale=1.2]{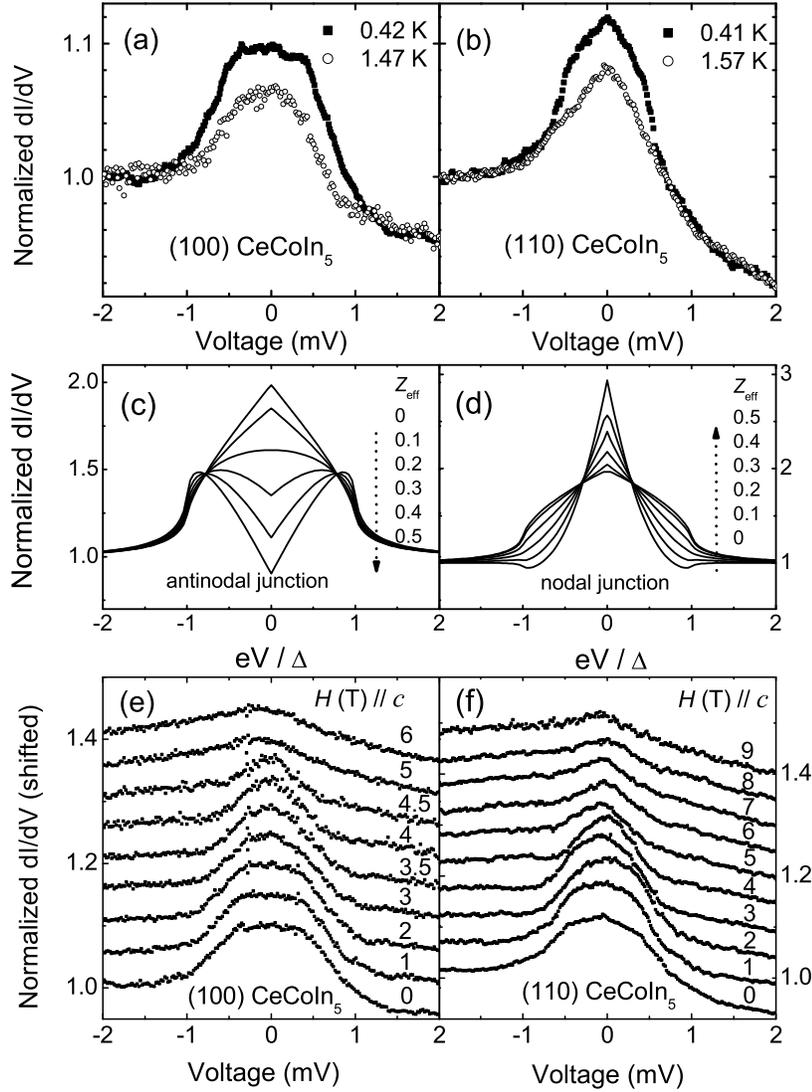}
\end{center}
\caption{\label{fig:dWevidence} (a) and (b) are normalized dI/dV of the (100) and (110) CeCoIn$_5$ junctions, respectively. (c) and (d) are normalized dI/dV curves calculated by the $d$-wave BTK formula ($\Gamma$=0 and $T$=0) for antinodal and nodal junctions, respectively. (e) and (f) are magnetic field (parallel to the $c$-axis) dependence for the (100) junction at 400 mK and the (110) junction at 420 mK, respectively. After \cite{park08prl}.}
\end{figure}

We now address the superconducting order parameter symmetry in CeCoIn$_5$. As seen in the simulation in figure~\ref{fig:dBTK}, a cusp-like shape is predicted for $d$-wave symmetry if there is no barrier ($Z_{\scriptsize \textrm{eff}} = 0$). The zero effective barrier strength condition is difficult, if not impossible, to achieve in conventional-metal/heavy-fermion-superconductor point-contact junctions because of some existing surface layer or even some effect of the Fermi velocity mismatch. We therefore assume $Z_{\scriptsize \textrm{eff}}$ is always finite. In this case, a cusp-like or peaked structure, even for small finite $Z_{\scriptsize \textrm{eff}}$, would never be observed in the antinodal direction. There always exist smearing effects due to finite temperature, finite quasiparticle lifetime and depairing. Thus, the interpretation of a flat conductance shape as we observed in (001) and (100) CeCoIn$_5$ junctions is ambiguous even between $s$- and $d$-wave symmetry. This was reported earlier for (001) junctions \cite{park05}, where a $d$-wave fit gives slightly better results. Therefore, it is essential to obtain conductance data in the $ab$-plane to determine the pairing symmetry. Figures~\ref{fig:dWevidence}(a) \& (b) compare such data along the (100) and (110) directions. While both spectra exhibit similar background asymmetry as seen in the (001) junction, their detailed shapes in the sub-gap region are quite different from each other. Namely, the (100) data look flat, similarly to the (001) data, whereas the (110) data are cusp-like. A comparison of these experimental observations with theoretically predicted behaviours gives strong evidence for $d$-wave symmetry, in fact $d_{x^2-y^2}$ \cite{park08prl}. Figures~\ref{fig:dWevidence}(c) \& (d) show calculated curves in the small-$Z_{\scriptsize \textrm{eff}}$ limit using the $d$-wave BTK model \cite{kashiwaya96}. The calculated conductance curves at $Z_{\scriptsize \textrm{eff}} = 0$ are identical, as discussed earlier. For finite $Z_{\scriptsize \textrm{eff}}$, they develop completely different features: double-peak and dip for the antinodal junction vs. narrower and shaper peak at zero bias for the nodal junction. We conclude that the (100) data are consistent with antinodal junction behaviour with $Z_{\scriptsize \textrm{eff}}\sim 0.3$ and the (110) data represent nodal junction behaviour with a smaller $Z_{\scriptsize \textrm{eff}}$. Therefore, the superconducting order parameter symmetry in CeCoIn$_5$ is $d_{x^2-y^2}$-wave, not $d_{xy}$-wave.

This spectroscopic result resolves the controversy of the node locations \cite{izawa01,aoki04,vorontsov06}. This definitive determination of the pairing symmetry narrows down possible candidates for the bosonic mode involved in the microscopic pairing mechanism \cite{scalapino87}. It is notable that AR measurements along different crystallographic directions provide spectroscopic evidence for the order parameter symmetry. This is possible via detecting the sign change albeit ABS are smeared to finite energy due to the higher junction transparency compared to tunnel junctions. Therefore, like ABS tunneling spectroscopy, PCS can provide phase-sensitive information in unconventional superconductors \cite{vanharlingen95,tsuei00}. As for the anisotropy in the order parameter amplitude, the conductance widths are similar for the (100) and (110) junctions probably due to the large tunneling cone effect. This is an issue to be addressed more rigorously in the future. A cusp-like feature similar to that in our data is observed in Goll {\it et al.}'s measurements into the $ab$-plane of CeCoIn$_5$ but the precise crystallographic orientation was not reported \cite{goll06}.

As discussed previously, ABS are known to split spontaneously or by applied magnetic field. This splitting occurs due to the Doppler shift effect in the bound state energy arising from a coupling of the quasiparticle velocity ($\vec{v}_{\scriptsize \textrm{F}}$) to the superfluid momentum ($\vec{P}_{\scriptsize \textrm{S}}$), $\vec{v}_{\scriptsize \textrm{F}} \cdot \vec{P}_{\scriptsize \textrm{S}}$. According to Tanaka {\it et al.} \cite{tanaka02}, the critical field for splitting of ABS is proportional to the junction transparency. Point-contact junctions generally have high transparency, so splitting of ABS would only be accomplished at very high field. This is completely consistent with our field-dependent measurements, as shown in figures~\ref{fig:dWevidence}(e) and (f). Other possibilities for the non-splitting of ABS have also been suggested, including the tunneling cone effect \cite{hentges04} and atomic scale disorder \cite{greene04} in cuprate tunnel junctions.

\section{How Andreev reflection can occur in heavy fermions}

Our conductance data on CeCoIn$_5$ is a clear example for AR in heavy fermions. Also, there have been many reports on the observation of AR in other heavy-fermion superconductors \cite{lohneysen96,naidyuk98,goll93,wilde94,goll95,wilde96,naidyuk96physb,obermair98}. As mentioned previously, this experimental observation is in conflict with what the BTK theory predicts as in (\ref{eq:barrierZ}): AR should be completely suppressed due to the large mismatch in the Fermi velocities of quasiparticles. Thus, how AR is possible in heavy fermions is a longstanding issue. Deutscher and Nozi{\'e}res \cite{deutscher94} addressed this issue and proposed a scenario that the velocity ($\bar{v}_{\scriptsize \textrm{F}}$) of injected or extracted particles are not the same as the quasiparticle velocity ($v_{\scriptsize \textrm{F}}$) as seen in specific heat or coherence length measurements. Namely,
\begin{equation}
v_{\scriptsize \textrm{F}} = \bar{v}_{\scriptsize \textrm{F}} \cdot z,
\label{dn}
\end{equation}
where $z$ is the mass renormalization factor and $1/z \equiv 1+\lambda$ is the mass enhancement factor \cite{eliashberg94}. In heavy fermions, $z$ is very small. Lukic considered generalized boundary conditions in the BTK problem and proposed that one should not interpret the parameter $Z_{\scriptsize \textrm{eff}}$ as representing materials properties but just as parameterization of the unknown microscopic details of the interface \cite{lukic}. Even though these arguments provide a way to explain the discrepancy between the BTK theory and experimental observations, they do not account for why AR is strongly reduced in heavy fermions. It is our conjecture that this issue might be intimately connected to why the conductance shows an asymmetry, so we focus on this issue in the next two sections.

\section{Conductance asymmetry}

In our PCS data on CeCoIn$_5$, there are two prominent and common features that are not seen in other class of superconductors: highly suppressed AR and background conductance asymmetry. It is noted that these features have also been observed in other reports on the PCS of heavy-fermion superconductors \cite{naidyuk98,goll06,goll06a,park08physb2}.

Although our qualitative analysis of the point-contact conductance spectra using the $d$-wave BTK model gives spectroscopic evidence for the superconducting order parameter symmetry in CeCoIn$_5$, it does not give us precise information on the gap amplitude. As discussed previously, existing models taking the Fermi surface mismatch and possible breakdown of Andreev approximation in heavy fermions into consideration do not  quantitatively account for our data. This is an indication that a crucial element is missing in the currently available models. This situation is in striking contrast to those cases in the other class of superconductors, where the BTK models have been so successfully applied \cite{naidyuk05,deutscher05}. We attribute the origin of this failure to the intrinsic properties of heavy fermions, more specifically, the coexistence of multiple electronic components \cite{nakatsuji04,curro04,yang08}.

\begin{figure}
\begin{center}
\includegraphics[scale=1.0]{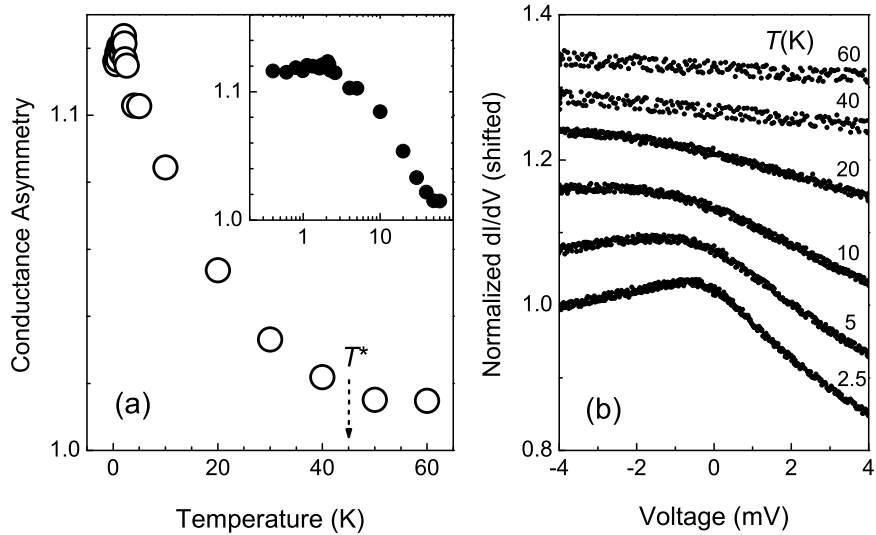}
\end{center}
\caption{\label{fig:asymmvsT} (a) Temperature dependence of the background conductance asymmetry, defined as the ratio between conductance values at --2 mV and at +2 mV from the (001) CeCoIn$_5$ data in figure~\ref{fig:PCS001}. $T^*$ is the heavy fermion coherence temperature. The inset is a semi-logarithmic plot of the same data. (b) Temperature-dependent evolution of the normal state conductance.  After \cite{park08prl}.}
\end{figure}

In figure~\ref{fig:asymmvsT}(a), it is found that the temperature dependence of the conductance asymmetry qualitatively follows that of the spectral weight for the coherent heavy electron liquid reported by Nakatsuji, Pines and Fisk \cite{nakatsuji04}. Also, there is a systematic evolution of this background conductance peak position as shown in figure~\ref{fig:asymmvsT}(b): it moves from the far-negative bias side towards zero bias with decreasing temperature. Yang and Pines \cite{yang08} recently reported a universal scaling behavior between various experimental data including our conductance asymmetry in figure~\ref{fig:asymmvsT}(a) and the spectral weight for the emergent heavy-electron liquid.

According to Harrison's theorem \cite{harrison61}, the electronic DOS of a simple metal cannot be measured using tunnel junctions since it cancels out the velocity factor in the conductance kernel. This theorem appears to hold in many cases where the single particle picture is valid. Applying this argument to point-contact junctions consisting of simple metals, the electronic DOS is not expected to be measured from such measurements. The conductance of a metallic contact is given by
\begin{equation}
\frac{dI}{dV}(V) \propto \int\int v N(E)\frac{\partial f(E - eV)}{\partial (eV)} dE d\Omega,
\label{mmjcn}
\end{equation}
where $v$ is the velocity, $N(E)$ is the electronic DOS of the counter-electrode, $f$ is the Fermi distribution function and $d\Omega$ is the differential solid angle \cite{naidyuk98}. Over a small bias region, the conductance is expected to be flat. Beyond that region, it is expected and observed to be curved downward due to increased scattering, say, with phonons, as the energy increases \cite{srikanth92}, and this shape is not attributed to the DOS. The downward curvature in metallic junctions is contrary to the upward curvature in tunnel junctions \cite{srikanth92}. Now, the question is whether Harrison's theorem still holds in the case where the many-body interaction is not negligible, as in heavy fermions.

The observed conductance asymmetry in our data on CeCoIn$_5$ might be strongly tied to this issue. As discussed in detail in \cite{park08lt25}, the conductance asymmetry cannot be explained by the models proposed in the literature. Certainly the asymmetry indicates it is easier to remove electrons from than to add to the heavy-fermion electrode CeCoIn$_5$. However, a simple argument that it is more difficult to add than to remove an electron from the f-orbital does not work because the f-electrons do not remain localized, but are itinerant, in heavy-fermion compounds. The argument that competing orders cause a conductance asymmetry \cite{hu06} doesn't appear relevant to our case because it is observed in all three members of the Ce-based 1-1-5 heavy fermions \cite{park08lt25}, no matter how far they are from the competing region. The persistency of conductance asymmetry in magnetic field up to 9 T (see figures~\ref{fig:dWevidence}(e) \& (f)) rules out a connection to the argument based on non-Fermi liquid behaviours \cite{shaginyan07}. In the past, the conductance asymmetry observed in PCS of heavy fermions was frequently attributed to the large Seebeck coefficients of heavy fermions \cite{itskovich85,naidyuk85,paulus85} combined with contact's being in the thermal regime. However, this model doesn't agree with our experimental data on 1-1-5 heavy fermions \cite{park08lt25}. A simple counter argument to this scenario is why the asymmetry persists in the superconducting state as seen in figure~\ref{fig:PCS001}(a), where the Seebeck effect disappears \cite{bel04} and, thus, the asymmetry should, too.

Anders and Gloos proposed that both the conductance asymmetry and reduced AR in heavy-fermion superconductors could be caused by the strongly energy-dependent quasiparticle scattering and DOS of heavy fermions \cite{anders97}. This theory does account for our observed conductance features but, since the calculations are done in a Green's function formalism, the microscopic physical mechanism is not obvious. Nowack and Klug \cite{nowack92} used Boltzmann transport equations to deal with electronic scattering by an energy-dependent DOS of heavy fermions. In this model, the asymmetry arises from the DOS centered at a finite energy. We note that the DOS effect of heavy fermions is invoked in both models.

\section{Fano interference effect and conductance modeling} 

We have proposed \cite{park08prl} a two-channel conductance model based on the two-fluid picture \cite{nakatsuji04} and originally assuming a Lorentzian form of the heavy electron DOS. Here, the two channels are into the superconducting heavy electron liquid and a possible normal conducting light electron liquid \cite{tanatar05}. It was assumed that the two channels are independent of each other. In theory it was argued that superconductivity should be induced to the light normal electron liquid due to the proximity effect \cite{barzykin07} but this issue is not yet settled. Our experimental data were found to accurately fit to this model at lowest temperatures but the data couldn't be fit with the model at higher temperatures; the fit failing more severely with increasing temperature. This is technically because the background conductance shape is not strictly Lorentzian. Instead, it is observed \cite{park08prl} to resemble a Fano line shape \cite{fano61}, implying that the two conductance channels need to be entangled instead of being independent. Motivated by our two-channel conductance model, Ara\'ujo and Sacramento recently formulated a BTK model adopting the two-fluid picture for heavy fermions \cite{araujo08}. In this model, AR is reduced essentially due to the existence of a normal conduction channel, similarly to our model, but they claim that the two channels should be allowed to mix via an interference term instead of being independent.

The Fano resonance \cite{fano61}, originally discovered in electron-helium inelastic scattering cross sections, has been frequently observed in a variety of condensed matter and other physical systems. In essence, it is a manifestation of an interference effect between waves that have passed through two different paths containing discrete and continuum states, respectively. This interference causes an asymmetry in the resonance spectra whose line shapes can be reproduced by the Fano formula \cite{fano61}. Recently, this phenomenon has been a subject of intensive investigations in quantum dot experiments \cite{kouwenhoven01,goldhaber-gordon98,cronenwett98,gores00,vanderwiel00,kobayashi02,kobayashi03} and STS of single adatoms and molecules \cite{madhavan98,li98,manoharan00,jamneala00,odom00,madhavan01,nagaoka02,knorr02,schneider02,heinrich04,wahl04,zhao05,neel07,wahl07,vitali08}. For example, Kobayashi {\it et al.} observed systematic evolution of the asymmetry in the differential conductance across a mesoscopic ring containing a quantum dot in one arm and a continuum in the other arm \cite{kobayashi02,kobayashi03}. Madhavan {\it et al.} investigated conductance behaviours of single adatoms deposited onto metallic substrates using STM \cite{madhavan98,madhavan01}. The measured differential conductance showed featureless shapes from non-magnetic atoms, whereas highly asymmetric curves from magnetic atoms. Their proposed conductance formula including the Fano resonance term successfully accounts for the measured data \cite{madhavan01}. Here, the interference occurs between tunneling electrons into the conduction band and into the local d orbital of an adatom.

\begin{figure}[t]
\begin{center}
\includegraphics[scale=1.1]{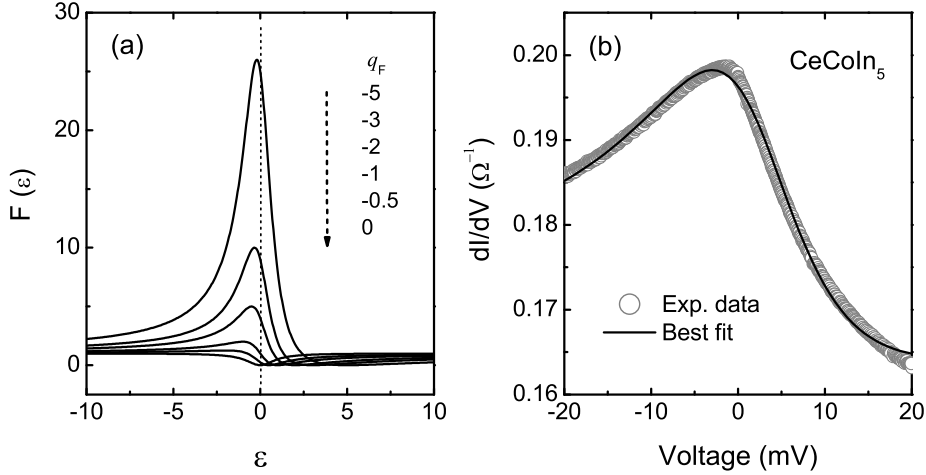}
\end{center}
\caption{\label{fig:Fanofit} (a) Fano line shapes as a function of the Fano parameter, $q_{\scriptsize \textrm{F}}$. (b) Fano fit to the data on (001) CeCoIn$_5$ at 2.43 K. Fitting parameters used are $q_{\scriptsize \textrm{F}}=-2.14$, $E_0=2.23$ meV, $\Lambda/2=11.13$ meV, $C$ = 0.0061 $\Omega^{-1}$, and $G_0$ = 0.164 $\Omega^{-1}$. After \cite{park08lt25}.} 
\end{figure}

The original Fano formula is given as:
\begin{equation}
F(\epsilon) = \frac{(q_{\scriptsize \textrm{F}}+\epsilon)^2}{1+{\epsilon}^2}
\label{eq:fano},
\end{equation}
where $q_{\scriptsize \textrm{F}}$ is the Fano parameter, $\epsilon \equiv (E-E_0)/ \frac{\Lambda}{2}$ and $E_0$ and $\Lambda$ are the resonance energy and width, respectively. Using this formula, Fano lines are simulated as a function of $q_{\scriptsize \textrm{F}}$, as plotted in figure~\ref{fig:Fanofit}(a). In the extreme limit of $|q_{\scriptsize \textrm{F}}| \to \infty$, the line becomes symmetric and represents a Lorentzian resonance. In the opposite limit, i.e., $|q_{\scriptsize \textrm{F}}| \to 0$, it corresponds to an antiresonance. In between these extremes, the line is asymmetric with a peak positioned at $\epsilon=1/q_{\scriptsize \textrm{F}}$.

Adopting a similar formula to that used by Madhavan {\it et al.} \cite{madhavan01}, we were able to fit the data and extract useful information \cite{park08lt25}. We set up a model that the differential conductance is determined by the following equation containing the Fano formula (\ref{eq:fano}) as the kernel. 
\begin{equation}
\frac{dI}{dV}(V) = C\int_{-\infty}^{\infty} dE \frac{\partial f(E-eV)}{\partial (eV)} F(\epsilon)+G_0,
\label{eq:fanocond}
\end{equation}
where $C$ is a prefactor for the Fano resonance contribution and $G_0$ is a constant conductance. An example of the fit using this formula is shown in figure~\ref{fig:Fanofit}. Here, the used fit parameters are: $q_{\scriptsize \textrm{F}}=-2.14$, $E_0=2.23$ meV, $\Lambda/2=11.13$ meV, $C$ = 0.0061 $\Omega^{-1}$, and $G_0$ = 0.164 $\Omega^{-1}$. In our previous paper \cite{park08prl}, we already reported a Fano resonance fit to the normalized conductance data over a narrower bias region using the following parameter values: $q_{\scriptsize \textrm{F}}=-1.9$, $E_0=0.55$ meV, $\Lambda/2=2.6$ meV, $C$ = 0.03646, and $G_0$ = 0.8458. From these two fits, it is clearly seen that the parameter values vary widely depending on the voltage range for fitting. Therefore, it is highly desirable to obtain and fit conductance data over a wide enough voltage range.

One of the unresolved issues in our previous fit using a Lorentzian DOS \cite{park08prl} was why the resonance energy is located below the Fermi level. Theoretically, the Kondo resonance in Ce-based Kondo lattice systems is known to occur at or above the Fermi level, which is also observed in photoemission measurements on several heavy-fermion systems \cite{reinert01} (cf. see \cite{wahl04} for the Kondo resonance energy position in single impurity STM experiments). In the current Fano fit, it is seen that our data can be reproduced by the Fano formula with the resonance energy above the Fermi level, in agreement with theoretical predictions and other experiments. This was possible with the choice of a negative value for $q_{\scriptsize \textrm{F}}$, which forces the resonance energy, $E_0$, to be positive. In general, $q_{\scriptsize \textrm{F}}$ is a complex number since the Fano resonance arises from an interference effect, as discussed above. Thus, the negative $q_{\scriptsize \textrm{F}}$ value we obtained can be understood as due to the phase factor in the interference. Similar analysis using negative $q_{\scriptsize \textrm{F}}$ was also reported in some single impurity STM experiments \cite{odom00}. Recently Yang \cite{yang08a} reported results from an extensive fit using a modified Fano formula to our previously reported data \cite{park08prl} over the $\pm$ 50 mV range at two temperatures (1.47 K and 20.62 K) and data over the $\pm$ 4 mV range at several temperatures between 2.5 K and 60 K. The calculated curves fit to the data well, but the implications of the extracted parameter values remain to be investigated as their values depend on the range of voltage used in the fitting, as demonstrated above. Also, the development of a microscopic model beyond phenomenological arguments is needed, as was done for the single impurity STS study \cite{madhavan01}. We discuss relevant issues in the following. 

One might raise questions about the validity of our analysis based on a Fano effect. First, how is it possible to observe a Fano resonance from point-contact (not tunneling) measurements and from the Kondo lattice systems? Regarding this issue, we note that there is an example \cite{neel07} where continuous evolution of the Fano resonance across a tunneling-metallic junction boundary is observed from STM measurements on single adatoms. This implies that it may also be possible in PCS. However, one should note a configurational difference between the Kondo lattice and single adatoms since in the latter the Fano (Kondo) resonance is widely known to be sensitive to the environment around the impurity atom. Second, what is the origin of the interference? In the two-fluid picture of heavy fermions \cite{nakatsuji04}, there are two components in the electronic spectra in CeCoIn$_5$: one is the heavy electron liquid and the other is conduction electrons that do not participate in the hybridization. Then, one can imagine that the interference occurs between these two channels \cite{park08prl,yang08a}. According to the theoretical derivation by Madhavan {\it et al.} \cite{madhavan01}, the Fano parameter $q_{\scriptsize \textrm{F}}$ is given by the ratio of two matrix elements: $q_{\scriptsize \textrm{F}}=A/B$, where $A$ is the coupling to a local orbital either direct or indirect via hybridization and $B$ is the coupling to the conduction electron continuum. It is to be investigated further which of the two terms for $A$ is dominant in PCS of the Kondo lattice system. As seen in the above fit, there exists a large background term, $G_0$, which might indicate that there exists substantial contribution from the channel into the conduction band without causing the interference. The Fano line shape in our data persists over a wide temperature range including both normal and superconducting states, implying that the interference must be persistent over a wide temperature range. Third, how can we combine a Fano resonance with AR or tunneling in heavy-fermion superconductors? As seen in figure~\ref{fig:PCS001}, the background conductance asymmetry affects the shape of the AR conductance. Thus, a successful microscopic theory should also explain this connection. In regards to this issue, we note that Flint {\it et al.} recently proposed a theory on the AR into a composite paired superconductor \cite{flint08}. They claimed that it is possible for electrons to cotunnel into the Kondo lattice and this would provide an enhanced contribution to AR with a Fano resonant structure. It is interesting to look into the similarities and differences between their Fano resonant structure and the asymmetric conductance shape in our model. 

\section{Remaining issues and future work}

{\it Boundary conditions in the BTK problem.} Although the theory by Deutscher and Nozi{\`e}res \cite{deutscher94} explains how AR is possible in heavy fermions, what role is played by the Fermi velocity mismatch is not clear. Related to this topic, Ara\'ujo and Sacramento \cite{araujo08} reported recently that the argument by Deutscher and Nozi{\`e}res can be proved explicitly by solving this problem for a normal-metal/heavy-fermion-superconductor interface. Quantitative measure of the Fermi velocity mismatch effect in PCS might be achieved by measuring AR conductance from a ballistic junction involving a pair with well-matched Fermi velocities, e.g., Au/Nb, but without any oxide layer formed at the interface.

{\it Tunneling cone effect.} In order to detect ABS unambiguously, making measurements in the tunneling limit will provide more flexible diagnostics. Thin film growth of CeCoIn$_5$ has been attempted but the quality is not yet good enough for all-thin-film based tunnel junctions. As an alternative, it is worthwhile trying to deposit an artificial tunnel barrier onto single crystals.

{\it AR in heavy-fermion-metal/conventional superconductor.} Despite some success reported in CeCoIn$_5$/Nb point contacts, the magnitude of AR signal in such junctions is not well determined due to the intervening oxide layer on the Nb tip \cite{park08physb1}. Further measurements on such junctions but without oxide interface layers will provide important clues to many issues on AR in heavy fermions. A related experiment will be exploring whether proximity effect exists in such junctions \cite{greene85,otop03}. The fact that AR is frequently observed in heavy fermions implies that proximity effect should exist because it is well known that AR is at the heart of this phenomenon \cite{klapwijk04}.

{\it Fano effect in Kondo lattice systems.} Although our proposed model has shown some evidence for this effect, it is not yet studied theoretically whether such effect could exist and, if so, is measurable by PCS. Also, it is an open question how to apply this picture to explain the reduced AR as well as the conductance asymmetry. Microscopic derivation of this effect may lead to further insightful investigations of the Kondo lattice physics. 

\section{Conclusions}

In conclusion, PCS on CeCoIn$_5$ has not only provided spectroscopic information on the superconducting order parameter but also clarified several experimental and theoretical issues. Sets of reproducible conductance spectra were obtained as a function of temperature, magnetic field and crystallographic orientation. A qualitative analysis of the data based on the extended BTK model shows that the order parameter in CeCoIn$_5$ has a $d_{x^2-y^2}$ symmetry. Detailed analysis of the conductance asymmetry led us to a phenomenological model based on a possible Fano effect in this Kondo lattice system. Further experimental and theoretical investigations of this model will lead us to more deepened understanding of the Kondo lattice physics. 

\ack
We greatly acknowledge longstanding collaborations with E. D. Bauer, J. L. Sarrao and J. D. Thompson at Los Alamos National Laboratory, C. Capan and Z. Fisk at the University of California-Irvine and P. C. Canfield at the Iowa State University. We are grateful to J. E. Elenewski, V. Lukic, A. J. Leggett, M. Stone, Y.-F. Yang, D. Pines, R. Flint and P. Coleman for theoretical discussions and P. W. Anderson for bringing up the issue of Harrison's theorem. This work was supported by the U.S. National Science Foundation under the Grant No. DMR 07-06013 and by the U.S. Department of Energy, Division of Materials Sciences under Award No. DEFG02-07ER46453, through the Frederick Seitz Materials Research Laboratory and the Center for Microanalysis of Materials at the University of Illinois at Urbana-Champaign.

\end{document}